\documentclass[10pt,aps,pr,showpacs,superscriptaddress,groupedaddress,nofootinbib]{revtex4}  

\usepackage{longtable}
\usepackage{ulem}   
\usepackage{comment} 
\usepackage{datetime}
\usepackage{tabularx}
\usepackage{float}
\restylefloat{table}
\usepackage[cmtip,arrow]{xy}
\usepackage{pb-diagram,pb-xy}

\makeatletter

\usepackage{amsmath}
\usepackage{amssymb}
\usepackage{amsthm}
\usepackage[pdftex]{color}
\usepackage[pdftex,colorlinks,citecolor=blue,linkcolor=blue,urlcolor=blue]{hyperref} 
\usepackage{graphicx}
\usepackage{dcolumn} 
\usepackage{bm} 

\usepackage{longtable}

\usepackage{ulem}   
\usepackage{comment} 
\usepackage{datetime}

\newtheorem{definition}{Definition}
\newtheorem{proposition}{Proposition}

\newtheorem{claim}{Claim}
\newtheorem{corollary}{Corollary}

\newcommand{\beq}{\begin{equation}}
\newcommand{\eeq}{\end{equation}}
\newcommand{\bea}{\begin{eqnarray}}
\newcommand{\eea}{\end{eqnarray}}
\newcommand{\barr}{\begin{array}}
\newcommand{\earr}{\end{array}}

\long\def\begincomment#1\endcomment{}

\newcommand{\cG}{\mathcal{G}}

\usepackage{stmaryrd}

\usepackage{pgf,tikz}
\usetikzlibrary{arrows}

\usepackage{bm}
\usepackage{bbold}

\usepackage{mathtools}

\DeclarePairedDelimiterX\braket[2]{\langle}{\rangle}{#1 \delimsize\vert #2}

\begin{document}


\title{Ward identity violation for melonic $T^4$-truncation}

\author{Vincent Lahoche} \email{vincent.lahoche@cea.fr}  
\affiliation{Commissariat \`a l'\'Energie Atomique (CEA, LIST), 8 Avenue de la Vauve, 91120 Palaiseau, France}

\author{Dine Ousmane Samary}
\email{dine.ousmanesamary@cipma.uac.bj}
\affiliation{Commissariat \`a l'\'Energie Atomique (CEA, LIST), 8 Avenue de la Vauve, 91120 Palaiseau, France}
\affiliation{Facult\'ee des Sciences et Techniques (ICMPA-UNESCO Chair), Universit\'e d'Abomey-Calavi, 072 BP 50, Benin}

\date{\today}

\begin{abstract}
Referring to  recent works concerning the functional renormalization group for tensorial group fields theories [Class. Quantum Grav. 35 (2018) 195006] and [Physical Review D 98, 126010 (2018)], this  paper gives in-depth explanation for, the ambiguity around the search of fixed points in the Wetterich flow equation and the Ward-Takahashi identities.  We consider the $U(1)$-tensors models  and discuss the non-existence of phase transition taking into account the Ward-identities as new constraint along the flow. We prove that  the quartic melonic tensorial  group fields theories without closure constraint are devoid with  the fixed points and therefore the phase transitions.
\end{abstract}


\maketitle

\section{Introduction} \label{sec1}
It is no longer  doubt to assert that the well defined theory of quantum gravity 
is the most challenging  question of modern physics which remains unsolved. Several approaches have been proposed  to tackle this issue. Our approach is called tensorial group field theory (TGFT) 
 see \cite{Lahoche:2018oeo}-\cite{BenGeloun:2012pu} and references therein: 
It is a class of fields theories, which aims at describing quantum theory of gravity  without  any geometric background for the definition of its fundamental degrees of freedom or dynamical equations. This theory is built around group fields theories (GFTs) 
\cite{Oriti:2010hg}-\cite{Oriti:2013jga} and tensors models (TMs) \cite{Rivasseau:2014ima}-\cite{Rivasseau:2013uca} and  allows to enjoy renormalization and asymptotic freedom in quite some generality \cite{BenGeloun:2012pu}, and for which we can show another case of its coexistence with a Wilson-Fisher fixed point \cite{Benedetti:2014qsa}.  The very  interesting point  concerning  the TM which needs to be discussed is the existence of a large $N$-limit dominated by the graphs called melons, thanks to the Gurau breakthrough discovered a few year ago \cite{Gurau:2011xq}-\cite{Gurau:2010ba}. More recently the so called Sachdev-Ye-Kitaev (SYK) quantum mechanics model which consists of $N$-Majorana fermions with random interactions  are showed to admit this same large $N$-limit. The melons  would certainly be the point of intersection of the physical reality and required a particular consideration: See \cite{Choudhury:2017tax}-\cite{Gurau:2016lzk} and references therein.

  TGFT has  recently  stimulated  efforts  in  two  closely
related  directions.  In the first, the functional renormalization group (FRG) is applied to several models and the analysis of the behaviors around  UV and  IR are given. The computation of the UV and IR fixed points and the phase diagrams allowed to prove the asymptotically freedom and safety \cite{Lahoche:2016xiq}-\cite{Geloun:2016qyb}.  Generally, the central tool of this study is based on the choice of  a truncation  i.e.  the average effective action $\Gamma_s$ and the  regulator $r_s$.  In  the  related  development, it was pointed out that there exists an alternative way to address the FRG by improving the choice of these approximations \cite{Lahoche:2018oeo}-\cite{Lahoche:2018vun}. Some class of new relations called structure equations, and the set of Ward-Takahashi (WT) identities are used  in the flow equations, which control not only the choice of the truncation but also the regulator. The WT-identities appear as new constraints on the flow and we will discuss in detail its influences in this paper.

Let's quickly recall some important results regarding the FRG for TGFT given in \cite{Lahoche:2018oeo} and \cite{Lahoche:2018vun}. First of all, in \cite{Lahoche:2018vun} the leading order melonic contributions is taken into account in our renormalization program. We showed that  it is always possible to address the issue of  FRG without choice of truncation and in appropriate way we get fixed points. This makes our method very encouraging and totally different from the usual FRG methods \cite{Lahoche:2016xiq}-\cite{Geloun:2016qyb}, in the sens that they can help to show the convergence in the flow and to identified the physical fixed points without doubt of it consistency. The same analysis is considered in \cite{Lahoche:2018oeo} by taking into account not  only the melonic sector, but also  other leading order contributions called pseudo-melons. The combinatorial analysis of these two sectors (melon and pseudo-melon) is considered. The flow of the couplings and mass are given in the symmetric phase. In the set of these two papers   the structure equations and WT-identities are considered  in our analysis. Surprisingly the WT-identity  enforced one new constraint on the anomalous dimension which  allows  a doubt on all fixed points obtained in a lot of recent works (at least in the TGFT models without closure constraint) \cite{Lahoche:2018oeo}-\cite{Lahoche:2018vun}, \cite{BenGeloun:2018ekd}-\cite{Geloun:2016qyb}. The aim of the following paper is to clarify this point in detail for quartic melonic interactions $T^4$. 

The paper is organized as follows: In section \eqref{secm} we provide briefly the usefull  definitions and  ingredients concerning the TGFT on the $U(1)$ group.  In section \eqref{sec2} the FRG is studied for quartic melonic tensor models. We derive analytically the corresponding fixed point and study its behaviors. The flow diagram is also given and we can easly  conclude in favor of the asymptotically freedom of these kind of  models  (see also \cite{Rivasseau:2015ova} for more detail). In section \eqref{sec3} the WT-identity and the structure equations are scrutinized in detail for the symmetric phase melonic sector. In  section \eqref{sec4} the violation of the WT-identity is identified for any choice of the regulator \eqref{regulator} and  for arbitrary dimensions $D$ and $d$. Section \eqref{sec5} is devoted to FRG without truncation. The structure equations derived in  section \eqref{sec3} are used to identify  the fixed points and compared with what we obtained in the case of ordinary truncation given in section \eqref{sec2}. Finally in section \eqref{sec6}  we provide the conclusion of our work.

\section{Tensorial group field theory}\label{secm}
TGFT is defined by the functional action $S[\phi,\bar \phi]$,  which depends on the fields $\phi$ and its conjugate $\bar\phi$,  taking values on $d$ copies of arbitrary group $\mathrm{G}$ of dimension $D$.
 \bea
\phi,\,\bar\phi:\, \mathrm{G}^d\rightarrow \mathbb{C}
\eea 
In the particular case we choose the Abelian group $\mathrm{G}:=U(1)^D$.  In the FRG point of view the classical action $S[\phi,\bar \phi]$ is replaced by the $s$-dependent  action $S[\phi,\bar \phi]+{\rm Tr}_G[\phi\, r_s \bar\phi]$ which taking into account the scale fluctuation when $s$ walks $\mathbb{R}$.   This modification introduces the IR regulator $r_s$ which satisfy the following boundary conditions
\bea\label{limreg}
\lim_{s\to -\infty }r_s=0,\quad \lim_{s\to \infty }r_s=\infty.
\eea
This implies that at the scale $s=\log \Lambda$ where $\Lambda$ is the UV cutoff,  the regulator $r_s$ remains very large and the average  effective action $\Gamma_s$  is reduced to the  classical action $S[\phi,\bar \phi]$. For the rest we consider only  the Fourier transform of the fields $\phi$ and $\bar\phi$ denoted respectively  by $T_{\vec p}$ and $\bar{T}_{\vec p}$,  $\vec p\in (\mathbb{Z}^{D})^d$ written as (for $\vec g\in U(1)^{Dd}$, $g_{j\ell}=e^{i\theta_{j\ell}}$):
\bea
\phi(\vec \theta\,)=\sum_{\vec p\in\mathbb{Z}^{Dd}}\,T_{\vec p} \,e^{i\sum_{j=1}^d\sum_{\ell=1}^D\theta_{j\ell}p_{j\ell}}\\
\bar\phi(\vec \theta\,)=\sum_{\vec p\in\mathbb{Z}^{Dd}}\,\bar T_{\vec p}\, e^{-i\sum_{j=1}^d\sum_{\ell=1}^D\theta_{j\ell}p_{j\ell}}.
\eea

  In the statistical mechanics point of view the field theory  is defined by the partition function:
\bea
Z_s[J,\bar J]=\int\,dT\,d\bar T e^{-S[T,\bar T]-\langle T, \,r_s \bar T\rangle+\langle J,\bar T\rangle+\langle T,\bar J\rangle},
\eea
where the scalar product $\langle\cdot,\cdot\rangle$ on $\mathbb{Z}^{Dd}$ is defined by: 
\bea
\langle a,b\rangle=\sum_{\vec p\in\mathbb{Z}^{Dd}}a_{\vec p}b_{\vec p}.
\eea
Recall that the action $S[T,\bar T]$ is splitted  as 
\bea
S[T,\bar T]=S_{\text{kin}}[T,\bar T]+S_{\text{int}}[T,\bar T].
\eea  
The kinetic term is  $S_{\text{kin}}[T,\bar T]=\langle T, \,C^{-1}\bar T\rangle$, with the propagator $C(\vec p\,)$ expressed in momentum space  by $C(\vec p\,)=(\vec p\,^{2\alpha} +m^{2\alpha})^{-1}$ and $\alpha$ a  parameter given by:
\bea\label{alpha}
\alpha=\frac{D(d-1)}{4}.
\eea 
The quartic interaction $S_{\text{int}}[T,\bar T]$ is suppose to be tensor invariant.
\bea
S_{\text{int}}[T,\bar T]\equiv\sum_{i=1}^d\,\lambda_i\,\vcenter{\hbox{\includegraphics[scale=0.7]{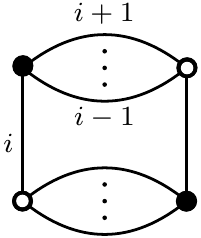} }}\,.\label{interaction4}
\eea
The model is said to be quartic and the interactions are melonic.
Because of the UV divergences, we introduce a regularization which suppress the high momenta contributions. Note that there are different choices of the regularization functions. In this note we consider only the dimensional regularization method.
 For this, we suppose that  the propagator is supported by the Gaussian measure with UV regularized propagator as:
\bea
C_{\Lambda}(\vec p\,)=\frac{\vartheta_\Lambda(\vec p\,^{2\alpha}/\Lambda^{2\alpha})}{\vec p\,^{2\alpha}+m^{2\alpha}}=\int\,d\mu_{C_\Lambda}\,T_{\vec p\,}\bar T_{\vec p\,},
\eea
with an arbitrary cut-off function $\vartheta_\Lambda$.

Now let us provide  some explanations about  the choice of the parameter $\alpha$.  We consider the  set of one vertex two point  graphs denoted by  $G({\mathfrak b})$ obtained from a connected tensorial bubble ${\mathfrak b}$,  having a maximum number of faces. The canonical dimension  $d_{\mathfrak b}$ of the bubble $\mathfrak b$  referring to \cite{Lahoche:2018oeo}-\cite{Lahoche:2018vun} is:
\bea
\tilde{d}_{\mathfrak b}=2\alpha -\max_{\cG\in G({\mathfrak b})}(\omega(\cG)).
\eea
Using the multiscale analysis and the power counting theorem, the divergent degree of the corresponding Feynman graph $\cG$ is 
\bea
\omega(\cG)=-2\alpha L(\cG)+DF(\cG),
\eea
where $L(\cG)$ is the number of lines and $F(\cG)$ the number of internal faces of $\cG$.  We can show that $\max(\omega(\cG))=-2\alpha+D(d-1)$.  The melons are caracterized by   $\tilde{d}_{\mathfrak b}=0$ and then  we recover the formula \eqref{alpha}.

The FRG we proposed here is based on the so called Wetterich equation \cite{Freire:1996db}-\cite{Wetterich:1989xg}:
\bea\label{Wetterich}
\partial_s \Gamma_s={\rm Tr}\,\partial_s r_s (\Gamma^{(2)}_s+r_s)^{-1},
\eea 
in which:  $\Gamma_s,\, -\infty<s<\infty$ is the average effective action which interpolates between the classical action $S$ (in the UV) and the full effective action $\Gamma$ (in the IR), such that this full effective action is obtained for the value $s\to -\infty$ i.e. in the IR limit.  $\Gamma^{(2)}_s$ is the second order partial derivative respect to the mean fields $M$ and $\bar M$ and $r_s$ is the IR regulator. The general form of this regulator is chosen to be
\bea\label{regulator}
r_s(\vec p\,)=Z k\,^{2\alpha} f\left(\frac{\vec p\,^{2\alpha}}{k^{2\alpha}}\right),\,\, k=e^s.
\eea
The function $ f\left(\frac{\vec p\,^{2\alpha}}{k^{2\alpha}}\right)$ is chosen such that the relation \eqref{limreg} is well satisfied. $Z$ is the wave function renormalization.
 When $s$ walks in $\mathbb{R}$ the flow equations enable us to interpolate smoothly between UV and IR phenomena. Using the Wetterich equation \eqref{Wetterich}, and by choosing the truncation and the regulator in appropriate way, it is easy to derived  the flow diagrams and  computed the fixed points.  In the case of quartic melonic tensor model the flow equations may be solved analytically and the fixed points can be given simply. 
The search of fixed points is important and essential in field theories especially in GFT.  They allow to understand if phase transitions exist and is probably the way to reconstruct our universe under a  geometrogenesis  scenario \cite{Mandrysz:2018sle}-\cite{Bonzom:2011zz}.  Unfortunately the violation of the WT-identity in the search of fixed points for several TGFT models without closure constraint studied in the litteratures  \cite{Lahoche:2018oeo}-\cite{Lahoche:2018vun}, \cite{BenGeloun:2018ekd}-\cite{Geloun:2016qyb}  allows us to conclude that no physical  fixed points can be found for these class of models. We will scrutinize   this statement in detail in section \eqref{sec4}.

\section{Solving flow equations: Truncations}\label{sec2}

There is no longer any doubt to recall that the Wetterich equation is
 generally non-solvable  exactly; and it is not a specificity of the TGFTs but a common situation in field theory. Extracting some informations on the non-perturbative equation \eqref{Wetterich} then requires approximations. In the TGFT context, the \textit{truncation method} and \textit{derivative expansion} have been largely investigated as a simple way to deal with the specific non-locality of the interactions. Basically, the strategy is to project the renormalization group flow into a reduced phase space of finite dimension. In the \textit{local potential approximation} discussed in the literature cited above (see Introduction), the truncations keep only the tensorial-invariant interactions, which are said to be \textit{locals}. All the truncation of the  effective actions are then of the form:
\bea
\Gamma[M,\bar{M}]= \Gamma_{\text{kin}}[M,\bar{M}]+\Gamma_{\text{int}}[M,\bar{M}]\,,
\eea
whose $\Gamma_{\text{int}}[M,\bar{M}]$ writes  as a \textit{finite sum} of connected tensorial invariants; schematically:
\begin{equation}
\Gamma_{\text{int}}[M,\bar{M}]=\vcenter{\hbox{\includegraphics[scale=0.8]{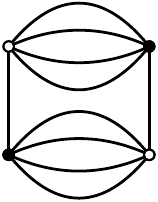} }}+\vcenter{\hbox{\includegraphics[scale=0.8]{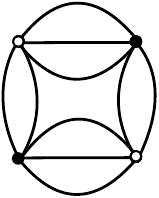} }}+\vcenter{\hbox{\includegraphics[scale=0.65]{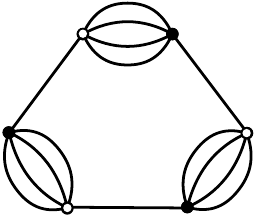} }}+\cdots\,.\\
\end{equation}
\medskip

\noindent
We expect that sufficiently far away from a fundamental scale $\Lambda$ the behavior of the renormalization group flow will be driven itself on a reduced phase space corresponding to renormalizable interactions; and a truncation around marginal interactions must be a not so bad approximation for the exact renormalization group flow. For the microscopic family of models discussed in the introduction, the minimally relevant truncation is then around quartic melonic interactions. In the same way, we have to keep only the relevant and marginal contributions in the derivative expansion of the kinetic action, respectively weighted with mass and wave-function renormalization:
\begin{equation}
\Gamma_{\text{kin}}[M,\bar{M}]=\int_{\mathrm{G}^d} d\textbb{g}d\textbb{g}^\prime \bar{M}(\textbb{g}^\prime)\left(-Z\Delta +m^2 \right)M(\textbb{g})\,,
\end{equation}
where $Z$ and $m^2$, as the effective action itself are expected to be dependent on the flow parameter $s$. The factor $Z$ may be easily cheeked to be marginal with respect to the Gaussian power counting. In full generality, a dependence of $Z$ on the means fields must be expected. However, for all applications considered in the literature, the authors set (explicitly or not) an expansion around vanishing means field. The flow is then built into a restrictive domain, called \textit{symmetric phase}, in which such an expansion makes sense. Among the characteristics of the symmetric phase, we can mention the following:
\begin{corollary}\label{cor1}
In the symmetric phase, the $2$-point effective function $G_{s}:=(\Gamma^{(2)}+r_s)^{-1}$ is diagonal : $G_{s,\vec{p},\vec{p}\,^\prime}=G_s(\vec{p}\,)\delta_{\vec{p},\vec{p}\,^{\prime}}$ and the functions which do not have the same number of derivatives with respect to $M$ and $\bar{M}$ vanish. 
\end{corollary}
\noindent
For an extensive discussion, the reader may consult  \cite{Lahoche:2018oeo}-\cite{Lahoche:2018vun}. In the symmetric phase, the wave function renormalization and the anomalous dimension $\eta$ may be defined without reference to any truncation as:
\begin{definition} \textbf{Anomalous dimension:}\label{def1}
In the symmetric phase, the wave function renormalization and the anomalous dimension are defined as: (the index $s$ is discarded in $\Gamma^{(2)}_s$ to make the notation simple)
\begin{equation} 
Z:= \frac{d}{d\vec{p}\,^2} \Gamma^{(2)}(\vec{p}=\vec{0}\,)\,,\qquad \eta:= \frac{1}{Z} \frac{\partial Z}{\partial s}\,.
\end{equation}
\end{definition}

\noindent
Note that in this definition we have took into account the corollary \ref{cor1}, and assumed that $\Gamma^{(2)}$ is a function of a single momentum $\vec{p}\in (\mathbb{Z}^D)^d$. Finally, we assign the same coupling constant $\lambda$ for each of the $d$ melonic interactions keeping in the truncation. The flow equations for the $\phi^4$-truncation in the local potential approximation may be derived from the exact functional flow equation \eqref{Wetterich}. Taking the successive functional derivatives with respect to $M$ and $\bar{M}$, we obtain an infinite hierarchical system of coupled equations, the equation for $\partial_s \Gamma^{(2)}$ involving $\Gamma^{(4)}$, the one for $\partial_s \Gamma^{(4)}$ involving $\Gamma^{(4)}$ and $\Gamma^{(6)}$, and so on and so for. The truncation cuts this infinite tower of coupled equations up to the quartic melonic interactions, enforcing that:
\begin{equation}
\Gamma^{(n)} \approx 0\,,\qquad \forall n>4\,.
\end{equation}
Moreover, we work into a restricted domain between the deep UV corresponding to some fundamental scale $\Lambda$, and the IR where $k=e^s$ vanish. Explicitly: $\Lambda\gg k\gg 0$. Then, all the irrelevant contributions in the large $k$ limit are discarded from the analysis. Let us for instance discuss the flow equation for $2$-point function, $\partial_s \Gamma^{(2)}$. Deriving twice the exact flow equation, we get schematically :
\begin{equation}
\partial_s \Gamma^{(2)}=-2\lambda\,\sum_{i=1}^d \left\{\vcenter{\hbox{ \includegraphics[scale=0.6]{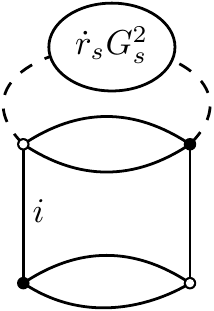}}}\,\, + \vcenter{\hbox{ \includegraphics[scale=0.6]{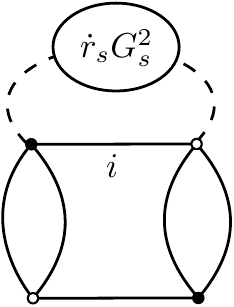}}}  \right\} \,,\label{flowtwo}
\end{equation}
the diagrams being computed with the effective propagator $\dot{r}_s G^2_s$, the ``dot" means that the derivative is  with respect to $s$. Explicitly:
\begin{equation}
\vcenter{\hbox{ \includegraphics[scale=0.6]{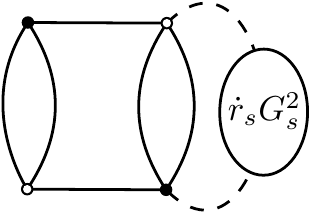}}} \,= \sum_{\vec{p}\in (\mathbb{Z}^D)^{(d-1)}} \frac{\dot{r}_s}{(Z\vec{p}\,^{2\alpha}+Zq^{2\alpha}+m^{2\alpha}+r_s)^2}
\end{equation}
where $q$ denotes the external momenta running through the effective loop. From an elementary power counting, the second contribution in the equation \eqref{flowtwo} must be discarded with respect to the first one from a relative scaling $k^{-D(d-2)}$. The relevant contributions correspond to the melonic ones, largely discussed in the random tensor models literature as the leading order diagrams on their $1/N$ expansion. In the TGFT context, melonic diagrams are recovered as the ones having the stronger relevant dependence on the cut-off, and with this respect appears naturally as the relevant ones in the large $k$ limit. As a result, only the melonic contractions are keeping in the computation of the flow equations. Finally, completing the definition \eqref{def1} with the definition of the \textit{effective mass} $m^{2\alpha}$ and \textit{effective melonic coupling} $\lambda$ at scale $s$ as:
\begin{equation}
m^{2\alpha}(s):=  \Gamma^{(2)}(\vec{p}=\vec{0}\,)\,,\qquad \lambda(s) := \frac{1}{4}\,\Gamma^{(4)}(\vec{0},\vec{0},\vec{0},\vec{0}\,)\,,\label{rencond}
\end{equation}
the leading order flow equations of  $m^2$, $Z$ and $\lambda$ is written as 
\begin{align}
\left\{
    \begin{array}{ll}
       \dot{m}^{2\alpha}&=-2d\lambda\, I_2(0)\,, \\
      \dot{Z}&= -2\lambda I_2^\prime(q=0)\,,\\
       \dot{\lambda}&=4\lambda^2 I_3(0)\,. \label{syst1}
    \end{array}
\right.
\end{align}
where $I_2^\prime:= \frac{d}{dq_1^{2\alpha}} I_2$, and we introduced the sums $I_n(q)$: 
\begin{equation}
I_n(q):= \sum_{\vec{p}\in (\mathbb{Z}^D)^{(d-1)}} \frac{\dot{r}_s}{(Z\vec{p}\,^{2\alpha}+Zq^{2\alpha}+m^{2\alpha}+r_s)^n}\,.
\end{equation}
The explicit computation of the sums $I_n$ requires to make a choice for the regulator function $r_s$. A common choice for standard field theories  and generally used in the TGFT context is the \textit{modified Litim's regulator} \cite{Litim:2001up}-\cite{Litim:2000ci}, which has been showed to be optimal in applications:
\begin{equation}
r_s(\vec{p}\,):=Z(e^{2\alpha s}-\vec{p}\,^{2\alpha})\theta(e^{2\alpha s}-\vec{p}\,^{2\alpha})\,,\label{Litim}
\end{equation}
where $\theta(x)$ stands for the Heaviside function.
A great interest for this regulator with respect to the other ones, except its optimal behavior, is that it allows to perform all the sums analytically for large $k$ using integral approximation:
\begin{equation}
 \sum_{\vec{p}\in (\mathbb{Z}^D)^{(d-1)}}  \to \int_{\mathbb{R}^{D(d-1)}}\,,
\end{equation}
and it is easy to see that all the sums required for the computations of the sums are all of the form:
\begin{equation}
J_{n}(R)=\int d^{D(d-1)}x  (\vec{x}\,^{2\alpha})^n\theta(R^{2\alpha}-\vec{x}\,^{2\alpha}) \,,\, n\in \mathbb{N}\,.
\end{equation}
The integral may be easily performed. Making a change of variable $y=x/R$ and using the integral representation for Heaviside function in term of  Dirac delta function as:
\begin{equation}
\theta(1-x)=\int_0^1dz \delta(z-x)\,,\qquad x\in\mathbb{R}^+\,,
\end{equation}
we get, up to the rescaling $y\to z^{1/2\alpha} y$:
\begin{equation}
J_{n}(R)=R^{2\alpha(2+n)}\,\iota(d,D)\int_0^1 dz z^{n+1} \,,\label{Jn}
\end{equation}
where $\iota(d,D)$ is defined as:
\begin{equation}
\iota(d,D):=\int d^{D(d-1)}y\, \delta(1-\vec{y}\,^{2\alpha})\,.
\end{equation}
The remaining integral over $y$ variables may be computed from the well known Feynman formula:
\begin{equation}
\prod_{i=1}^k\frac{1}{A_i^\beta}=\frac{\Gamma (k\beta)}{[\Gamma(\beta)]^k} \int_0^1 \prod_i du_i \delta\left(1-\sum_iu_i\right)\frac{\prod_iu_i^{\beta-1}}{(\sum_iA_iu_i)^{ k}}\,.
\end{equation}
Indeed, setting $A_i=1\,\forall i$, $k=D(d-1)$ and $\beta=1/2\alpha$, we get:
\begin{equation}
\iota(d,D)=2^{D(d-1)} \left[\Gamma\left(\frac{2+D(d-1)}{D(d-1)}\right)\right]^{D(d-1)}\,.
\end{equation}
The explicit expression for $I_n(0)$ and $I_n^\prime(0)$ may be easily deduced in term of $J_n$. Setting $R=e^s$, we obtain:
\begin{equation}
I_n(0)=Z\frac{\eta(R^{2\alpha}J_0(R)-J_1(R))+2\alpha R^{2\alpha} J_0(R)}{(ZR^{2\alpha}+m^{2\alpha})^n}\,, \label{In}
\end{equation}
and:
\begin{equation}
I_n^\prime(0)=Z\frac{\eta(J_1^\prime(R)-J_0(R))-R^{2\alpha}(\eta+2\alpha)J_0^\prime(R)}{(ZR^{2\alpha}+m^{2\alpha})^n}\,. \label{In2}
\end{equation}
The system \eqref{syst1} is not suitable to get some  non-trivial fixed points, because of their explicit dependence with respect on the parameter $s$. In order to obtain a closed and autonomous system of differential equations having non-trivial fixed points, we have to introduce the renormalized and dimensionless mass and couplings, defined as follow:
\begin{definition}
In accordance with their respective Gaussian canonical dimensions, we define the renormalized and dimensionless mass and coupling as:
\begin{equation}
\lambda=:Z^2\bar{\lambda}\,,\qquad m^{2\alpha}=:Ze^{2\alpha s} \bar{m}^{2\alpha s}\,.
\end{equation}
\end{definition}
Computing the sums from equations \eqref{In} and \eqref{In2}, and using the explicit expression for $J_n(R)$ given by equation \eqref{Jn}, we deduce the desired system which only involves dimensionless and renormalized parameters:
\begin{proposition}
In the intermediate UV sector $\Lambda\gg k\gg 1$, the truncated flow equations for renormalized and dimensionless essential and marginal parameters are written as:
\begin{align}
\left\{
    \begin{array}{ll}
       \beta_m&=-(2\alpha+\eta)\bar{m}^{2\alpha}-2\alpha d\bar{\lambda}\,\frac{\iota(d,D)}{(1+\bar{m}^{2\alpha})^2}\,\left(1+\frac{\eta}{6\alpha}\right)\,, \\
       \beta_{\lambda}&=-2\eta \bar{\lambda}+4\alpha\bar{\lambda}^2 \,\frac{\iota(d,D)}{(1+\bar{m}^{2\alpha})^3}\,\left(1+\frac{\eta}{6\alpha}\right)\,, \label{syst2}
    \end{array}
\right.
\end{align}
where $\beta_m:= \dot{\bar{m}}^{2\alpha}$,  $\beta_{\lambda}:=\dot{\bar{\lambda}}$ and:
\begin{equation}
\eta:=\frac{4\alpha\bar{\lambda} \iota(d,D)}{(1+\bar{m}^{2\alpha})^2-\bar{\lambda}\iota(d,D)}\,.\label{etatruncated}
\end{equation}
\end{proposition}
The proof of \eqref{syst2} and \eqref{etatruncated} are straightforward from the system \eqref{syst1}. Note that the  system \eqref{syst2} has two kinds of singularities. The first one is an  singularity line of equation $\bar{m}^{2\alpha}=-1$, coming from the denominators of the flow equations for mass and melonic coupling. The second singularity arise from the anomalous dimension denominators, and corresponds to a line of singularity, with equation:
\begin{equation}
\Omega(\bar{\lambda},\bar{m}^{2\alpha}):=(1+\bar{m}^{2\alpha})^2-\bar{\lambda}\iota(d,D)=0\,.
\end{equation}
This line of singularity splits the two dimensional phase space of the truncated theory into two connected regions characterized by the sign of the function $\Omega$. The region $I$, connected to the Gaussian fixed point for $\Omega>0$ and the region $II$ for $\Omega<0$. For $\Omega=0$, the flow becomes ill defined. The existence of this singularity is a common feature for expansions around vanishing means field, and the region $I$ may be viewed as the domain of validity of the expansion in the symmetric phase. Note that to ensure the positivity of the effective action, the melonic coupling must be positive as well. Therefore, we expect that the physical region of the reduced phase space correspond to the region $\lambda\geq 0$. From definition of the connected region $I$ and because of the explicit expression \eqref{etatruncated}, we deduce that :
\begin{equation}
\eta \geq 0\,,\qquad \text{In the symmetric phase}\,.
\end{equation}

\begin{figure}[H]
\begin{center}
\includegraphics[scale=0.27]{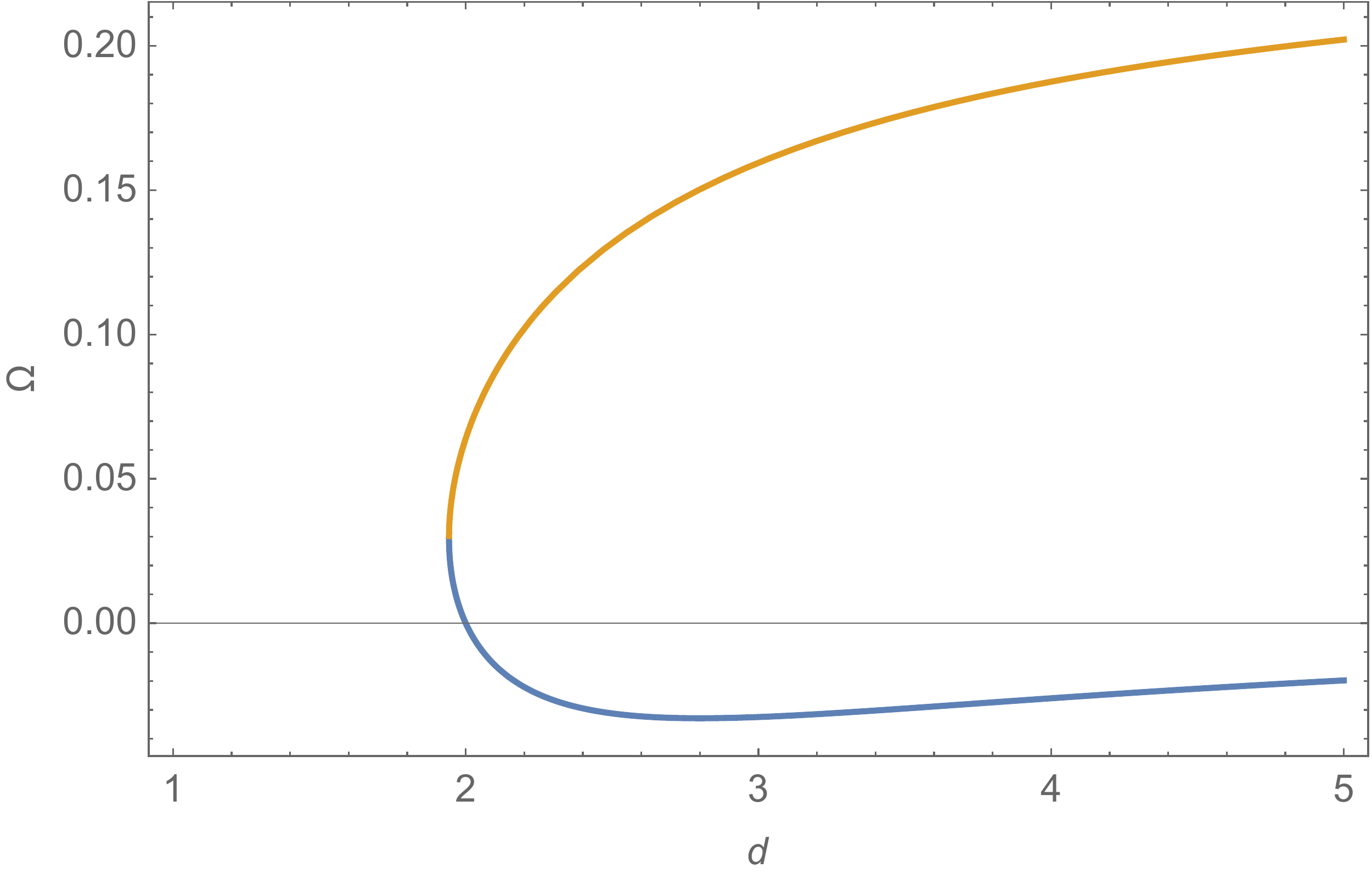} 
\caption{$\Omega(p_+)$ (in brown) and $\Omega(p_-)$ (in blue) in function of $d$. $\Omega(p_+)$ is a positive constant, going to $0.25$ when $d\to\infty$. In contrast, $\Omega(p_-)$ is negative, and goes to zero. }\label{fig1}
\end{center}
\end{figure}

\noindent
It is well known that these equations admit two non trivial fixed points, which can be explicitly computed :
\begin{equation}
p_\pm=\left\{
    \begin{array}{ll}
      m^{2\alpha}_\pm=-\frac{1+9d\mp\sqrt{\Delta}}{6+12d}\,, \\
\\
      \bar{\lambda}_\pm=\frac{-31\mp23\sqrt{\Delta}+3d(38+3d\pm\sqrt{\Delta})}{18(1+2d)^3\iota(d,D)}\,, \label{fp}
    \end{array}
\right.
\end{equation}
where $\Delta:=1+9(d-2)d$. Interestingly for $d>2$, it is easy to cheek that only $p_{+}$ is in the connected region $I$ ($\Omega(p_+)>0\,\forall d$) whereas $p_-$ is in the connected region $II$ ($\Omega(p_-)<0\,\forall d$). Figure \eqref{fig1} provides us a graphical illustration. 

The fixed point $p_+$ appears has an IR fixed point, having the same characteristic as the Wilson--Fisher fixed point: one relevant and one irrelevant direction. The integral curve of the relevant directions build a \textit{critical line} splitting the connected region $I$ into two regions. In the upper one all the RG trajectories go to the Gaussian fixed point, whereas in the lower one, all the RG trajectories escapes from the Gaussian region. 

\begin{figure}[H]\begin{center}
\includegraphics[scale=0.55]{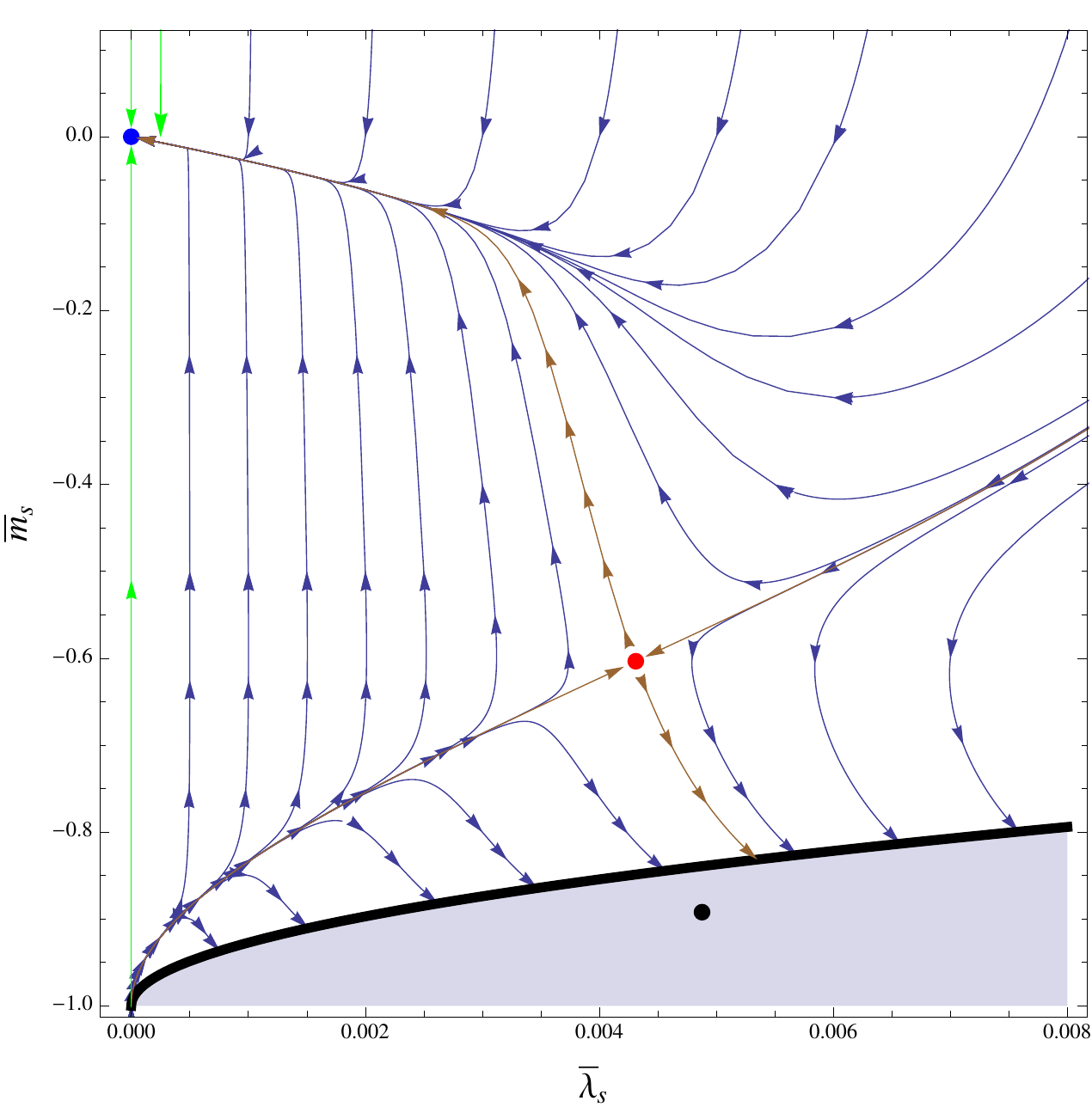} 
\caption{Numerical integration of the flow equations. The red and black points correspond to the non-Gaussian fixed points, the red one being into the connected region $I$. The critical line separating the region $I$ into two regions is pictured in brown and corresponds to the relevant direction (In the UV) of the red fixed point. The grey region corresponds to the region $II$, bounded with the line of equation $\Omega=0$, in black.}\label{fig2}\end{center}
\end{figure}

As pointed out in the literature, this situation involves a phase transition from a symmetric toward a non-symmetric phase. A specificity of this theory, pointed out in \cite{Benedetti:2015yaa}, is the coexistence of two exceptional situations: The existence of a non-trivial Wilson-Fisher fixed point and the asymptotic freedom in the vicinity of the Gaussian fixed point. To cheek this property, we may expand the previous flow equations \eqref{syst2} at the leading order in $\bar{\lambda}$, leading to:
\begin{equation}
\beta_\lambda\approx - \eta\bar{\lambda}\,,\qquad \eta\approx 4\alpha\bar{\lambda} \iota(d,D)\,. \label{asymptoticfreedom}
\end{equation}
Note that the minus sign comes from the exact equality between the one-loop vertex correction and the one-loop wave function correction. Moreover, this result does not depends on the choice of the truncation and regulating function, because of universality theorem. It has been pointed out as a common property of quartic melonic models  \cite{Rivasseau:2015ova}, but as we will discuss in the next section, it is a consequence of the Ward identities. The Figure \ref{fig2} provides a picture of the phase space structure obtained from a numerical integration for $D=1$ and $d=5$. For an extensive discussion, the reader may be consult the references \cite{Lahoche:2018vun},\cite{Benedetti:2015yaa}-\cite{Benedetti:2014qsa}.\\

\noindent
This scenario, referring a phase transition controlled from an IR fixed point has been largely considered in the literature as an universal property of the tensorial field theories. It was considered in particular as a solid argument in favor of a condensation scenario in GFT cosmological approaches (see \cite{Oriti:2015qva} and references therein).\\

\noindent
Before  move on to the next section, we have to make an important comment on the regularization scheme that we use. In the introduction we defined the regularized propagator $C_\Lambda$ with an arbitrary cut-off function $\vartheta_\Lambda$. Remark that,  we refered to some fundamental scale $\Lambda$ for the physical discussion. However, it is suitable for the rest of the paper to remove any appearance of this fundamental scale, assuming the continuum limit $\Lambda\to\infty$. This limit considerably simplify the derivation of the Ward identities, involving the derivative of the inverse propagator, $C_\Lambda^{-1}$ with respect to one of the momenta. Indeed, such a derivative includes contributions coming from the derivative of the regularization function, that we call ‘‘boundary contributions". These contributions have been considered in \cite{Lahoche:2018oeo}, and are irrelevant for our considerations. Because of the regulator $\dot{r}_s$, the flow equations are divergent free, end then insensitive on this limit. However, Ward identities introduce a divergent loop integration of all the momentum slices, which is finite only because of the counter-terms coming from the fact that the initial theory is just-renormalizable. Note that with this respect, the continuum limit makes sens only because the theory is asymptotically free, and has no Landau pole. To circumvent the divergence arising from the effective loop integration in the Ward identity without the additional subtleties coming from boundary terms, we use  a specific dimensional regularization, already discussed in \cite{Lahoche:2018oeo} for tensor field theories. From fixed $(D,d,\alpha)$, we allow a variation on the values of $D$ and  $d$ and  such that $\alpha$ remains fixed. Let $D'=D+\epsilon$ is now the group dimension. All the relevant amplitudes may be analytically continued in $D'$, and the divergences for the value $D'=D$ appear as pole in $1/\epsilon$, all of them being removed with counter-terms for renormalizability theorem. We will use of dimensional renormalization for convenience. Other regulators have been discussed in \cite{Carrozza:2017vkz}-\cite{Carrozza:2016tih}, in which  the authors argued that this boundary term does not depends on the running scale $k$ in the large $\Lambda$ limit, so that it disappears when we differentiate the Ward identity with respect to $k$, i.e. what we will make in order to extract a constraint on the RG flow.

\section{Ward--Takahashi identities and structure equations} \label{sec3}
It is well known that continuous symmetries in quantum field theories provides some non-trivial relations between Green's function, generally called \textit{Ward--Takahashi identities}. For TGFTs, such a relations comes from the specific unitary invariance of their interactions. Let $\textbf{U}=(U_1,\cdots,U_d)\in\mathbb{U}_{\infty}^{\times d}$, where $U_i\in\mathbb{U}_{\infty}$ are unitary matrices of infinite size, which we  denote by  $U_{i,\,p_1p_1^\prime}$ with $p_1,p_1^\prime \in \mathbb{Z}^D$. The transformation law for a tensor field $T$ is then:
\begin{equation}
\textbf{U}[T]_{p_1,\cdots,p_d}:=\sum_{\{p_i^\prime\}} \left(\prod_{j=1}^d \,U_{j,p_jp_j^\prime}\right)T_{p_1^\prime,\cdots,p_d^\prime}\,,
\end{equation}
where we used Einstein convention for summation of repeated indices. As a consequence of the tensorial invariance, the interacting part of the action $S_{\text{int}}$, involving power of field higher than $2$ satisfies: $\textbf{U}[S_{\text{int}}]=S_{\text{int}}\,\forall\, \textbf{U}$. In contrast, the kinetic action and the source terms are both non-invariant, respectively due to the $\vec{p}\,^2$ coming from the Laplacian and the sources, $J$ and $\bar{J}$. However, a stronger invariance comes from the translation invariance of the formal Lebesgue integration measure in the path integral defining as $Z(J,\bar{J})$: $\textbf{U}[Z]=Z$. The same argument have to be true for the $2$-point correlation function $\langle T_{\vec{p}}\bar{T}_{\vec{q}}\rangle$ (see \cite{Samary:2014tja} for more detail). Because of the translation invariance of the integral, we expect that this can be transformed as a trivial representation of the product $(\mathbb{U}_{\infty}^{\times d})\otimes (\mathbb{U}_{\infty}^{\ast\times d})$, the $\ast$ meaning complex conjugation. Translating this property for an infinitesimal anti-hermitian variation:
\begin{equation}
\textbf{U}=(\mathbb{I}+\epsilon_1,\mathbb{I},\cdots,\mathbb{I})=:\mathbf{1}+\vec{\epsilon}_1\,,
\end{equation}
where $\mathbb{I}$ denotes the identity transformation, we get:
\begin{equation}
\int \vec{\epsilon}_1[d\mu_{C_s}] T_{\vec{p}}\bar{T}_{\vec{q}} \,e^{-S_{\text{int}}}+\int d\mu_{C_s} \vec{\epsilon}_1[T_{\vec{p}}\bar{T}_{\vec{q}}] e^{-S_{\text{int}}}=0\,,\label{wardbrut}
\end{equation}
$d\mu_{C_s}:=dTd\bar{T}\,e^{-S_{\text{kin}}}$ being the Gaussian measure, and $C_s^{-1}:=C_{\infty}^{-1}+r_s$, where $C_{\infty}^{-1}(\vec{p}\,):=Z_{-\infty}\vec{p}\,^{2\alpha}+m_{\infty}^{2\alpha}$ denote the kinetic kernel including wave function and mass counter-terms. 
The second variation may be easily computed, straightforwardly:
\begin{align}
 \nonumber\vec{\epsilon}_i[\bar{T}_{\vec{p}}T_{\vec{q}}]&=-\epsilon_{p_ip_i^\prime}^*\bar{T}_{\vec{p}\,^\prime}\prod_{j\neq i} \delta_{p_jp_j^\prime}T_{\vec{q}}+\bar{T}_{\vec{p}}\, \epsilon_{q_iq_i^\prime}T_{\vec{q}\,^\prime}\prod_{j\neq i} \delta_{q_jq_j^\prime}\\\nonumber
&= \, \epsilon_{q_iq_i^\prime}\bar{T}_{\vec{p}}\prod_{j\neq i} \delta_{q_jq_j^\prime}T_{\vec{q}\,^\prime}-\epsilon_{p_i^\prime p_i}\bar{T}_{\vec{p}\,^\prime}\prod_{j\neq i} \delta_{p_jp_j^\prime}T_{\vec{q}}\\
&=\bar{T}_{\vec{p}}T_{\vec{q}_{\bot_i}\cup \{q_i^\prime\}} \epsilon_{q_iq_i^\prime}-\bar{T}_{\vec{p}_{\bot_i}\cup \{p_i^\prime\}}T_{\vec{q}}\,\epsilon_{p_i^\prime p_i}\,,
\end{align}
where $\vec{q}_{\bot_i}:=\vec{q}\,/q_i$. The variation of the second integral then becomes:
\begin{equation}
\int d\mu_{C_s} \vec{\epsilon}_1[T_{\vec{p}_1}\bar{T}_{\vec{p}_2}] e^{-S_{\text{int}}}=\delta_{\vec{p}_{\bot_1}\vec{q}_{\bot_1}}[G_s(\vec{p}\,)-G_s(\vec{q}\,)]\epsilon_{1,q_1p_1}\,,\label{var1}
\end{equation}
where we used corollary \ref{cor1}. The variation of the Gaussian measure, follows the same strategy, leads to:
\begin{equation}
\vec{\epsilon}_i[d\mu_{C_s}]=-\vec{\epsilon}_i\left[\sum_{\vec{p}}\bar{T}_{\vec{p}}\,C^{-1}_s(\vec{p}\,)T_{\vec{p}}\,\right]d\mu_{C_s}\,.
\end{equation}
The variation of the bracket follows the same strategy as the variation of $\bar{T}_{\vec{p}}T_{\vec{q}}$, and we get:
\begin{align}
\vec{\epsilon}_i[d\mu_{C_s}]=\sum_{\vec{p},\vec{q}}\epsilon_{q_ip_i}\delta_{\vec{p}_{\bot_i}\vec{q}_{\bot_i}}[C_s^{-1}(\vec{q}\,)-C_s^{-1}(\vec{p}\,)]\bar{T}_{\vec{p}}T_{\vec{q}}\,.\label{var2}
\end{align}
Using \eqref{var1} and \eqref{var2} in equation \eqref{wardbrut}, we get a non-trivial relation between $4$ and $2$-point functions:
\begin{align}
\nonumber\sum_{\vec{r}_{\bot_i},\vec{s}_{\bot_i}}&\delta_{\vec{r}_{\bot_i}\vec{s}_{\bot_i}}[C_s^{-1}(\vec{r}\,)-C_s^{-1}(\vec{s}\,)]\langle\bar{T}_{\vec{r}}T_{\vec{s}}\bar{T}_{\vec{p}}T_{\vec{q}}\rangle\\
&\quad=-\left[\delta_{\vec{p}_{\bot_i}\vec{q}_{\bot_i}}(G_s(\vec{p}\,)-G_s(\vec{q}\,))\delta_{r_iq_i}\delta_{s_ip_i}\right]\,.
\end{align}
The relation may be conveniently written in term of the effective vertex $\Gamma^{(4)}$, linking to the $4$-point function $\langle\bar{T}_{\vec{r}}T_{\vec{s}}\bar{T}_{\vec{p}}T_{\vec{q}}\rangle$ as:
\begin{equation}
\langle\bar{T}_{\vec{r}}T_{\vec{s}}\bar{T}_{\vec{p}}T_{\vec{q}}\rangle=\left(\Gamma_{\vec{r},\vec{s},\vec{p},\vec{q}}^{(4)}\,G_s(\vec{p}\,)G_s(\vec{q}\,)+\delta_{\vec{r}\vec{p}}\,\delta_{\vec{s}\vec{q}}\right)G_s(\vec{r}\,)G_s(\vec{s}\,)\,,\label{decomp}
\end{equation}
where we extracted the one-particle reducible parts of the $4$-point function. In the deep UV, for large $k$, a continuous approximation for variables is suitable. Then, setting $r_1=p_1$, $\vec{p}\to \vec{q}$, $r_1\to s_1$, we get finally:
\begin{proposition}\label{propWard}
In the deep UV, the $4$ and $2$-point functions are related as (on both sides, $r_1=p_1$):
\begin{align}\label{WT-id}
\sum_{\vec{r}_{\bot_1}} G_s^2(\vec{r}\,)\frac{dC_s^{-1}}{dr_1^{2\alpha}}(\vec{r}\,)\Gamma_{\vec{r},\vec{r},\vec{p},\vec{p}}^{(4)}=\frac{d}{dp_1^{2\alpha}}\left(C_\infty^{-1}(\vec{p}\,)-\Gamma^{(2)}(\vec{p}\,)\right)\,.
\end{align}
\end{proposition}
Note that, in the decomposition \eqref{decomp} we assumed, following corollary \ref{cor1} that all odd vertex functions vanish. The proposition \ref{propWard} involve the $4$-point vertex function, and as announced an effective loop which may be divergent. To give  more explanation  about the structure of this equation, we have to specify the structure of the vertex function. To this end, we use  this loop to discard the irrelevant contributions, and we keep only the \textit{melonic contribution} of the function $\Gamma^{(4)}$, say $\Gamma_{\text{melo}}^{(4)}$. In the symmetric phase, we define it as:
\begin{definition}
In the symmetric phase, the melonic contribution $\Gamma_{\text{melo}}^{(4)}$ may be defined as the part of the function $\Gamma^{(4)}$ which decomposes as a sum of melonic diagrams in the perturbative expansion.
\end{definition}

The structure of the melonic diagrams has been extensively discussed in the literature \cite{Lahoche:2016xiq}-\cite{Rivasseau:2013uca}, and specifically for the approach that we propose here in \cite{Lahoche:2018vun}. Formally, they are defined as the graphs optimizing the power counting; and they family can be build from the recursive definition of the vacuum melonic diagrams, from the cutting of some internal edges. Among there interesting properties, these construction imply the following statement:
\begin{proposition}\label{propmelons}
Let $\mathcal{G}_N$  be a $2N$-point 1PI melonic diagrams build with more than one vertices for a purely quartic melonic model. We call external vertices the vertices hooked to at least one external edge of $\mathcal{G}_N$ has :

\item $\bullet$  two external edges per external vertices, sharing $d-1$ external faces of length one. 
\item $\bullet$ $N$ external faces of the same color running through the interior of the diagram. 
\end{proposition}
As a direct consequence of the proposition \ref{propmelons}, we expect that melonic $4$-points functions is decomposed as:
\begin{equation}
\Gamma_{\text{melo}}^{(4)}=\sum_{i=1}^d \Gamma_{\text{melo}}^{(4),i}\,,
\end{equation}
the index $i$ running from $1$ to $d$ corresponding to the color of the $2$ internal faces running through the interiors of the diagrams building $ \Gamma_{\text{melo}}^{(4),i}$. Moreover, we expect that these monocolored components have the following structure:
\begin{equation}
\Gamma_{\text{melo}\vec{p}_1,\vec{p}_2,\vec{p}_3,\vec{p}_4}^{(4),i} = \vcenter{\hbox{\includegraphics[scale=0.5]{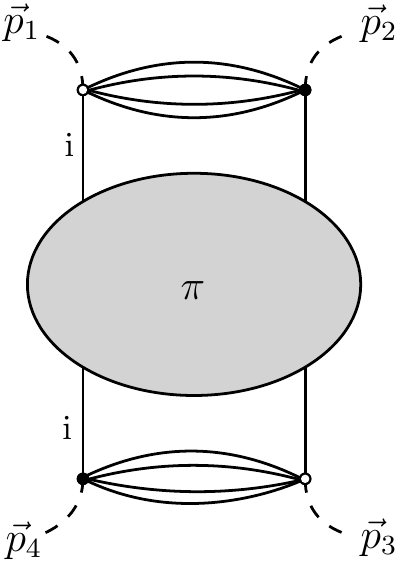} }}+ \vcenter{\hbox{\includegraphics[scale=0.5]{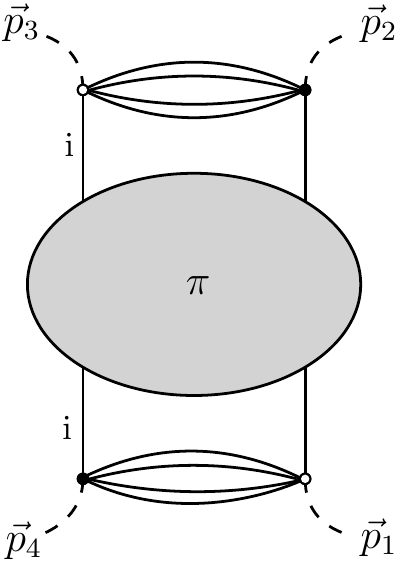} }}\,,\label{decomp4}
\end{equation}
the permutation of the external momenta $\vec{p}_1$ and $\vec{p}_3$ coming from Wick's theorem: There are four way to hook the external fields on the external vertices (two per type of field). Moreover, the simultaneous permutation of the black and white fields provides exactly the same diagram, and we count twice each configurations pictured on the previous equation. This additional factor $2$ is included in the definition of the matrix $\pi$, whose entries depend on the components $i$ of the external momenta running on the boundaries of the external faces of colors $i$, connecting together the end vertices of the diagrams building $\pi$.  \\

\noindent
Inserting \eqref{decomp4} into the Ward identity given from proposition \ref{propWard}, we get some contributions on the left hand side, the only one relevant of them in the deep UV being, graphically:
\begin{equation}\label{diage}
\vcenter{\hbox{\includegraphics[scale=0.5]{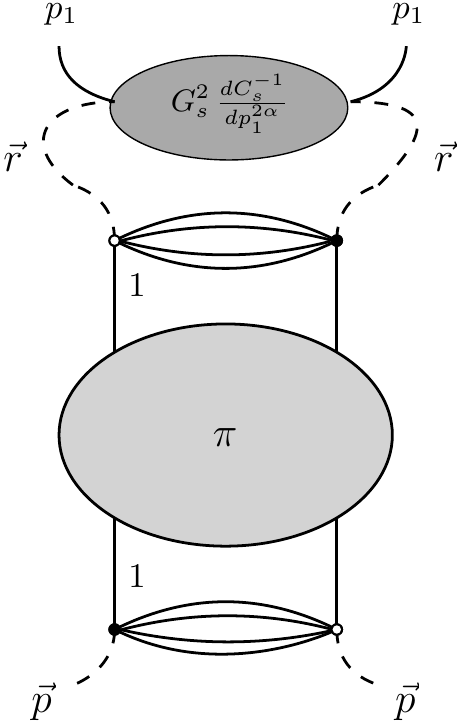} }}+\mathcal{O}(1/k)=\frac{d}{dp_1^{2\alpha}}\left(C_s^{-1}(\vec{p}\,)-\Gamma^{(2)}(\vec{p}\,)\right)\,.
\end{equation}
Setting $\vec{p}=\vec{0}$, and using the definition \ref{def1} as well as the definition of $C_\infty^{-1}$, the right hand side reduces to $Z_{-\infty}-Z$. Moreover, translating the diagram on the left hand side in the  equation \eqref{diage}, we get the equality:
\begin{equation}
Z_{-\infty} \mathcal{L}_s\, \pi_{00}=Z_{-\infty}-Z\,,
\end{equation}
where we have defined $Z_{-\infty} \mathcal{L}_s$ as:
\begin{equation}
Z_{-\infty} \mathcal{L}_s:=\sum_{\vec{p}\in(\mathbb{Z}^D)^{d}} \left(Z_{-\infty}+\frac{\partial r_s}{\partial p_1^{2\alpha}}(\vec{p}\,)\right)G^2_s(\vec{p}\,)\delta_{p_10}\,.
\end{equation}
Finally, from definition \eqref{decomp4} we expect that $\Gamma^{(4)}_{\text{melo},\vec{0},\vec{0},\vec{0},\vec{0}}=2\pi_{00}$, and because of the renormalization conditions \eqref{rencond} we must have the relation: $\pi_{00}=2\lambda(s)$, which ends  the proof of the following corollary:
\begin{corollary}
In the deep UV regime, the Ward identity between $4$ and $2$ point functions provides a non trivial relation between effective coupling and wave function renormalization:
\begin{equation}
2Z_{-\infty} \mathcal{L}_s\, \lambda=Z_{-\infty}-Z\,.\label{Wardutile}
\end{equation}
\end{corollary}
\noindent
In the melonic sector, this relation may be completed with a strong \textit{structure equation}, linking together the end point $\lambda_r:=\lambda(s=-\infty)$ and the effective coupling at scale $s$, $\lambda(s)$. We recall the main steps of the complete proofs, given in \cite{Lahoche:2018vun}. Let us denote by $Z_\lambda \lambda_r$ the bar coupling of the classical action, and $-4Z_\lambda \lambda_r \Pi$ the contributions of the perturbative expansion of $2\pi_{00}$ involving more than one vertex, such that :
\begin{equation}
2\pi_{00}=:4Z_\lambda \lambda_r (1+\Pi)\,. \label{defpi00}
\end{equation}
Because of the proposition \ref{propmelons}, we expect the following structure:
\begin{equation}
-4Z_\lambda \lambda_r \Pi=\vcenter{\hbox{\includegraphics[scale=0.6]{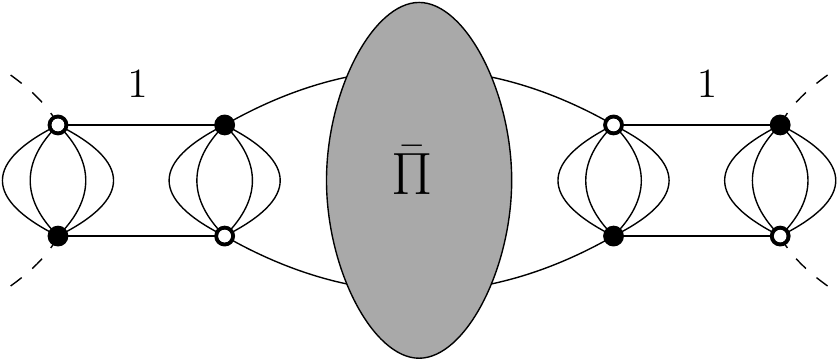} }}\,,
\end{equation}
where the grey bubble labeled with $\bar{\Pi}$ is a sum of Feynman graphs. Now, let us denote as $\bar{\Pi}^\prime$ the connected 1PI contribution to $\bar{\Pi}$, extracting from it disconnected contributions and external effective propagators. We get the equality:
\begin{equation}
-4Z_\lambda \lambda_r \Pi=\vcenter{\hbox{\includegraphics[scale=0.7]{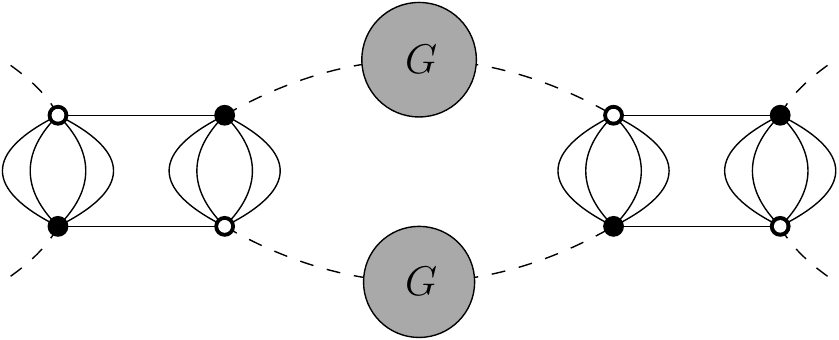} }}+ \vcenter{\hbox{\includegraphics[scale=0.7]{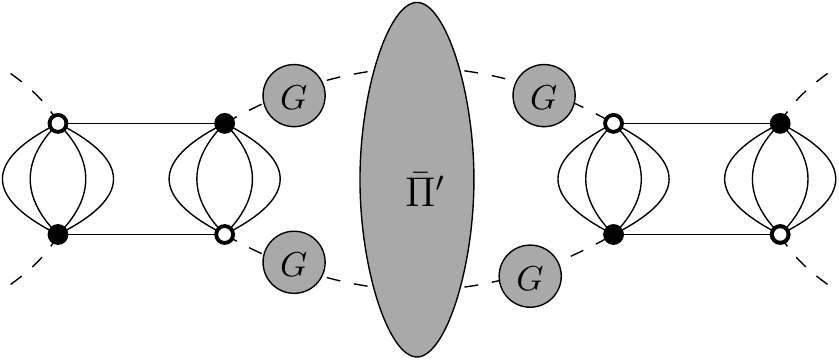} }}\,.\label{step1}
\end{equation}
Note that $\bar{\Pi}^\prime$ is of order $Z_\lambda\lambda_r$. Isolating this first order term, the argument leading to the equation \eqref{step1} may be repeated, and we deduce a closed equation for $\bar{\Pi}^\prime$:
\begin{equation}
\vcenter{\hbox{\includegraphics[scale=0.55]{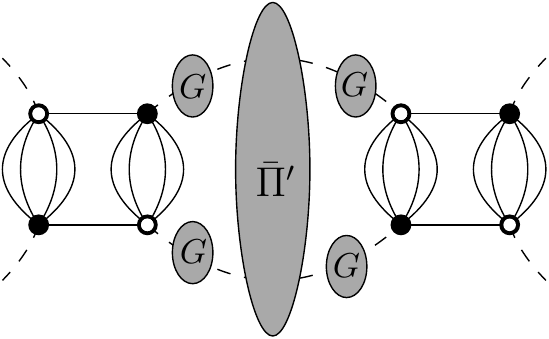} }}=\vcenter{\hbox{\includegraphics[scale=0.55]{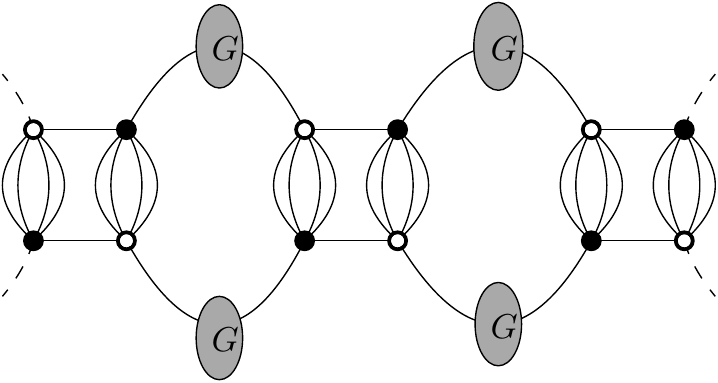} }}+\vcenter{\hbox{\includegraphics[scale=0.55]{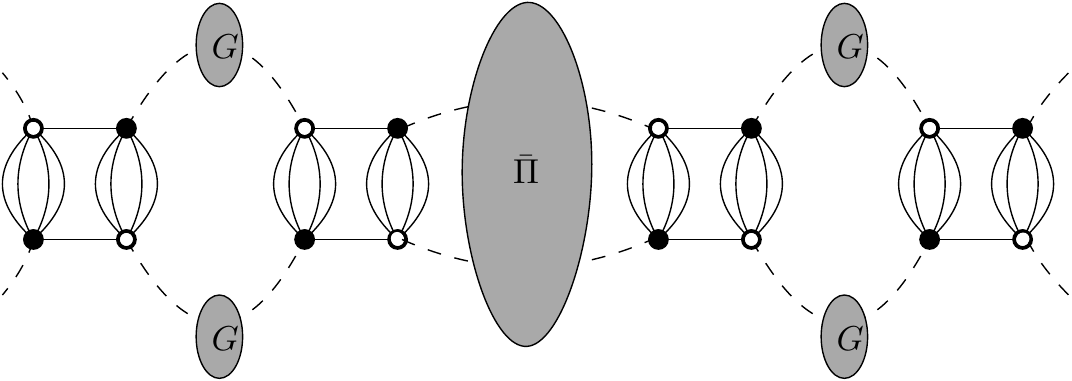} }}\,,
\end{equation}
which can be solved recursively as an infinite sum:
\begin{equation}
-4Z_\lambda \lambda_r \Pi=\vcenter{\hbox{\includegraphics[scale=0.6]{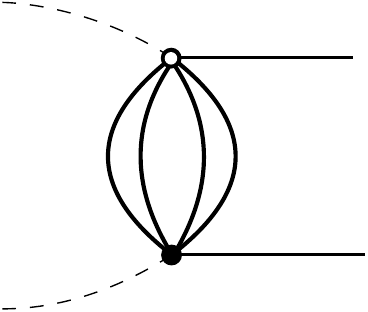} }} \left\{\sum_{n=1}^\infty \left(\vcenter{\hbox{\includegraphics[scale=0.6]{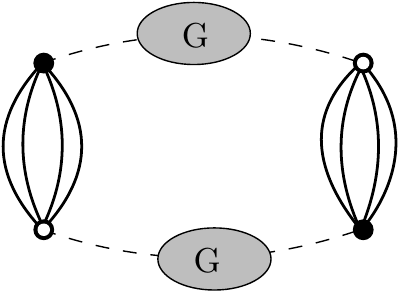} }}\right)^n\right\} \vcenter{\hbox{\includegraphics[scale=0.6]{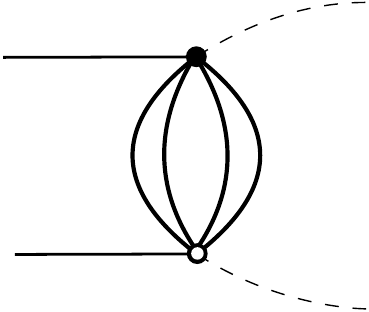} }}\,.\label{formalsum}
\end{equation}
The interior loop diagram $\vcenter{\hbox{\includegraphics[scale=0.3]{pimiddle.pdf} }}$ may be straightforwardly computed using Wick theorem for effective propagators $G_s$. We get:
\begin{equation}
\vcenter{\hbox{\includegraphics[scale=0.58]{pimiddle.pdf} }}=-2Z_\lambda \lambda_r \mathcal{A}_s\,,
\end{equation}
where we defined the quantity $\mathcal{A}_s$ as:
\begin{equation}
\mathcal{A}_s:=\sum_{\vec{p}\in(\mathbb{Z}^D)^{(d-1)}}\,G_s^2(\vec{p}\,)\,.
\end{equation}
Inserting the formal sum \eqref{formalsum} into the equation \eqref{defpi00}, we get:
\begin{equation}
\pi_{00}=\frac{2Z_\lambda \lambda_r}{1+2Z_\lambda \lambda_r \mathcal{A}_s}\,.
\end{equation}
The counter-term $Z_\lambda$ may be fixed from the boundary renormalization condition for $s\to-\infty$: 
\begin{equation}
\Gamma^{(4)}_{\text{melo},\vec{0},\vec{0},\vec{0},\vec{0}}\,(s=-\infty)=4\lambda_r\,,
\end{equation}
leading to:
\begin{equation}
Z_\lambda^{-1}=1-2\lambda_r \mathcal{A}_{s=-\infty}\,.
\end{equation}
Finally, because of the definition $\pi_{00}=2\lambda(s)$, we deduce the following important result:
\begin{proposition} \textbf{Structure equation for effective coupling:}
In the deep UV, the effective melonic coupling is given in term of the renormalized coupling $\lambda_r$ and the renormalized effective loop $\bar{\mathcal{A}}_s:=\mathcal{A}_s-\mathcal{A}_{s=-\infty}$ as:
\begin{equation}
\lambda(s)=\frac{\lambda_r}{1+2\lambda_r\bar{\mathcal{A}}_s}\,.\label{structure}
\end{equation}
\end{proposition}
With the help of this relation, the Ward identity \eqref{Wardutile} may be translated into a local version along the RG flow, hat we will discuss at the beginning of the next section.

\section{Ward identity violation}\label{sec4}
In this section we argue in favor of a strong violation of the Ward identities at the fixed point $p_+$ discussed in the section \ref{sec2}. In particular, we will show that, with the common hypothesis retain for truncation method, no fixed point have to exist in the symmetric phase, bounded with the singularity line $\Omega=0$ and $\lambda=0$. We refers to this restricted domain as $I^\prime$. \\

\noindent
Differentiating the equation \eqref{Wardutile} with respect to $s$ term by term, we get:
\begin{equation}\label{dotZ}
\dot{Z}=-2Z_{-\infty}\mathcal{L}_s \dot{\lambda}-2Z_{-\infty} \dot{\mathcal{A}}_s\,\lambda-2Z_{-\infty} \dot{\Delta}_s\,\lambda\,,
\end{equation}
where
\bea
\mathcal{L}_s:=\mathcal{A}_s+\Delta_s.
\eea
Deriving the equation \eqref{structure} with respect to $s$, we obtain a relation between $\dot{\mathcal{A}}_s$, $\lambda$ and $\dot{\lambda}$:
\begin{equation}
\dot{\lambda}=-2\lambda^2\,\dot{\mathcal{A}}_s\,,
\end{equation}
so that the previous equation \eqref{dotZ} becomes:
\begin{equation}
\dot{Z}=(Z_{-\infty}-2\lambda Z_{-\infty}\mathcal{L}_s)\frac{ \dot{\lambda}}{\lambda}-2Z_{-\infty} \dot{\Delta}_s\,\lambda\,.\label{label}
\end{equation}
The  computation of $ \dot{\Delta}_s$ requires to be carefully explored. We recall that $Z_{-\infty} \Delta_s$ includes a derivative of the regulating function $r_s$ with respect to  its momenta variables. For the class of regulating functions discussed in this paper see equation  \eqref{regulator}, $r_s$ is proportional to $Z$. In a first time we assume that $r_s$ is the modified Litim regulator \eqref{Litim}. For the computation of the flow equation in section \ref{sec2}, we make an assumption of the form of the kinetic effective action. Note that, in the symmetric phase, this assumption is the same like the previous works  \cite{Lahoche:2018oeo}-\cite{Lahoche:2018vun}, and we  impose the following approximation for the $2$-point function:
\begin{equation}
\Gamma^{(2)}(\vec{p}\,)=Z(s)\vec{p}\,^{2\alpha}+m^{2\alpha}(s)\,.\label{gamma2approx}
\end{equation}
Note that for the computation of the flow equations, this approximation has to be justified only in the restricted windows of momenta allowed by the distribution $\dot{r}_s$, and standard applications of the FRG support this approximation. However, it has been showed in  \cite{Lahoche:2018oeo}-\cite{Lahoche:2018vun} that using this approximation so far from this restricted domain, for the computation of $\mathcal{A}_s$ for instance, without use precaution, leads to a spurious conclusion for the sign of the universal one-loop beta function. For the Litim regulator, $r_s$, $\dot{r}_s$ and $\partial r_s/\partial p_1^{2\eta}$ provide the same windows of momenta: $\vec{p}\,^{2\alpha}\leq e^{2\alpha s}$. Therefore, if the approximation \eqref{gamma2approx} is suitable for the computation of the beta functions, it has to be a well approximation for the computation of $\Delta_s$. Explicitly we get:
\begin{equation}
Z_{-\infty} {\Delta}_s=-\frac{1}{Z}\frac{1}{2}\frac{\iota(d,D)}{(1+\bar{m}^{2\alpha})^2}\,,
\end{equation}
and the equation \eqref{label} becomes:
\begin{equation}
\dot{Z}=Z\frac{ \dot{\lambda}}{\lambda}-\frac{\lambda}{Z} \frac{\iota(d,D)}{(1+\bar{m}^{2\alpha})^2}\left(\frac{\dot{Z}}{Z}+\frac{2\beta_m}{1+\bar{m}^{2\alpha}}\,\right),
\end{equation}
where we have used the equation \eqref{Wardutile}. Finally, introducing the anomalous dimension to express the derivatives $\dot{Z}$, and using the definition of the dimensionless and renormalized coupling $\bar{\lambda}$, we deduce the following important result:
\begin{proposition}\label{mainprop}
In the deep UV, the Ward identity may be translated as a constraint between the beta functions and the anomalous dimension. In the symmetric phase, with the Litim regulator and keeping only the leading order of the derivative expansion, this constraint writes as:
\begin{equation}
\beta_\lambda=-\eta\bar\lambda\,\frac{\Omega(\bar{\lambda},\bar{m}^{2\alpha})}{(1+\bar{m}^{2\alpha})^2}+\frac{2\bar{\lambda}^2\iota(d,D)}{(1+\bar{m}^{2\alpha})^3}\beta_m\,.\label{eqmainprop}
\end{equation}
\end{proposition}
\noindent
Note that at first order, we recover the equation \eqref{asymptoticfreedom}; highlighting the deep nature of the compensation between vertex correction and wave function renormalization at one loop order. 
At a fixed point in the region $I^\prime$, by replacing equation \eqref{etatruncated} in \eqref{eqmainprop}, we get:
\bea
\frac{4\alpha\bar\lambda^2\iota(d,D)}{(1+\bar{m}^{2\alpha})^2}=0,\label{eqmainprop}
\eea
which admits only one solution: $\lambda=0$. Moreover, it is easy to cheek that the only fixed point having vanishing coupling has necessarily vanishing mass, and then match with the Gaussian fixed point. We then deduce the important statement:
\begin{corollary}\label{maincor}
The only fixed point compatible with the Ward identity in the region $I^\prime$ is the Gaussian fixed point.
\end{corollary}

The constraint given from proposition \ref{mainprop} have to be solved simultaneously
with the RG flow equation, projecting it locally into a physical phase space of dimension $1$, and the previous corollary show that there are no fixed point in this physical subspace:
\begin{claim}\label{claim1}
In the deep UV regime, the Litim regularized melonic--$T^4$--truncated RG flow has no physical IR fixed point in the symmetric phase.
\end{claim}
A violation of the Ward identities for relevant and marginal operators is expected to be an hard pathology. Indeed, viewing the RG flow like a mapping $\Gamma_k\to\Gamma_{k^\prime}$ we naively expect that the Ward identity constraint may be identically transported along the flow. The previous equations show that this expectation is wrong. Indeed, using the truncated values for the beta functions, and inserting them into the constraint \eqref{eqmainprop}, we get two curves $\bar{\lambda}_\pm(\bar{m}^{2\alpha})$, respectively is the region $I$ and in the region $II$, and it is easy to cheek that nothing happens when the RG trajectories cross the line. As a result, even if the initial conditions are chosen to enforce the constraint, any step of the RG flow transport the theory into an non-physic region. The origin of the problem may be expected to come from the fact that we don't solve exactly the RG flow; the approximation scheme introducing a spurious dependence on the regulation and an artificial violation of the Ward identity. However, in any cases there are no way to discard the Ward constraint, and the RG flow have to be completed with him. Moreover, it is easy to cheek that the same conclusion hold for any regulator $r_s$. Indeed, let us consider the general form of regulator in equation \eqref{regulator}. The equation for $\dot{Z}$ writes explicitly as:
\begin{equation}
\dot{Z}=4\lambda Z \sum_{\vec{p}\in\mathbb{Z}^{D(d-1)}} {\dot{r}_s}G^3_s(\vec{p}\,)-A\,,\label{equationZpoint}
\end{equation}
with:
\begin{equation}
A:=2\lambda \sum_{\vec{p}\in\mathbb{Z}^{D(d-1)}}\left[G^2_s(\vec{p}\,)\dot{r}_s^\prime-2\dot{r}_sr_s^\prime G_s^3(\vec{p}\,)\right]\,,
\end{equation}
and $r_s^\prime:=\partial r_s/\partial p_1^{2\alpha}$. In the same way:
\begin{equation}
2\lambda Z_{-\infty} \dot{\Delta}=A-4\lambda \sum_{\vec{p}\in\mathbb{Z}^{D(d-1)}} r_s^\prime\left(\,\dot{m}^{2\alpha}+Z\eta \vec{p}\,^{2\alpha}\right)G_s^3(\vec{p}\,)\,.
\end{equation}
Using equation \eqref{equationZpoint} to express $A$ in term of $\dot{Z}$, it is suitable to write $\bar{x}:=e^{d_x s} Z^{k_x} x$ for dimensionless quantities, $d_x$ being the canonical dimension of $x$, and $Z^{k_x}$ extracting its explicit dependence on $Z$. With this notation, we may write:
\begin{equation}\label{expression}
\frac{2\lambda Z_{-\infty} \dot{\Delta}}{Z}=-\eta-4\bar{\lambda}\bar{C}\,,
\end{equation}
where:
\begin{equation}
\bar{C}:=(2\alpha+\eta)\sum_{\vec{p}\in\mathbb{Z}^{D(d-1)}} G^3_s(\vec{p}\,)\left[\Big(\frac{\vec{p}\,^{2\alpha}}{e^{2\alpha s}}+\bar{m}^{2\alpha}\Big) \bar{r}_s^\prime-\bar{r}_s\right]\,.
\end{equation}
Note that to deduce this equation, we used of the elementary relation
\begin{equation}
\dot{r}_s:=(2\alpha+\eta)r_s(\vec{p}\,)-2\alpha Z\vec{p}\,^{2\alpha}f^\prime(\vec{p}\,^{2\alpha}/k^{2\alpha})\,,\label{equationderiv}
\end{equation}
easily to cheek from \eqref{regulator}. Now, let us move on to the denominator of $\eta$. The equation of the singularity line may be easily obtain as for  the Litim regulator, isolating on both sides of the equation \eqref{equationZpoint} the terms involving $\eta$. From equation \eqref{equationderiv}, and up to some straightforward manipulations, we get:
\begin{equation}
\Omega_f=1+2\bar{\lambda} \frac{\bar{C}}{2\alpha+\eta}-2\bar{\lambda} \sum_{\vec{p}\in\mathbb{Z}^{D(d-1)}}\,r_s(1+r_s^\prime)G^3_s(\vec{p}\,)\,,
\end{equation}
the index $f$ referring on the arbitrariness of the regulating function $f$. Replacing the expression \eqref{expression} into the equation \eqref{label}, we get:
\begin{equation}
\eta=\frac{\beta_\lambda+2\eta\bar{\lambda}}{\bar{\lambda}}+\eta+4\bar{\lambda}\bar{C}\,.
\end{equation}
Reaching a fixed point, the beta function vanish from definition $\beta_\lambda=0$, leading to the constraint:
\begin{equation}
X:=\eta+2\bar{\lambda} \bar{C}=0\,.
\end{equation}
Isolating $\bar{C}$ in term of the positive function $\Omega_f$ in the region $I^\prime$, we get finally:
\bea
X&=&(2\alpha+\eta)\left(\Omega_f+2\bar{\lambda} \sum_{\vec{p}\in\mathbb{Z}^{D(d-1)}}\,r_s(1+r_s^\prime)G^3_s(\vec{p}\,)\right)-2\alpha\,.
\eea
Note that the positivity of $\eta$ in the region $I^\prime$ does not depends on the choice of $f$. Moreover, the second term in the parenthesis on the right hand side must be positive defined in $I^\prime$, as it can be easily cheeked for standard choices of function $f$. As a result: 
\begin{equation}\label{boundX}
X\geq \eta \,,
\end{equation}
for any non-Gaussian fixed point in the region $I^\prime$. Therefore, for any non-Gaussian fixed point $\eta$ have to be a strictly positive quantity, the claim \ref{claim1} may be extended for any regulator as:
\begin{claim}\label{claim2}
For any choice of regulating function, the region $I^\prime$ is empty of non-Gaussian fixed point in the deep UV.
\end{claim}
\section{Beyond truncation method}\label{sec5}

One expect that higher truncations in the melonic sector does not improve the previous conclusions: The singularity line only depends on the quartic interactions, and the equation \eqref{eqmainprop} does not change. Moreover, we expect that our bound \eqref{boundX} discard any non-Gaussian fixed point in the region $I^\prime$ as well. Indeed, we obtain that so far from the deep UV regime, interactions beyond marginal sector becomes irrelevant, so that the positivity of the action ensures the positivity of the quartic melonic coupling $\lambda$, and therefore the positivity of the anomalous dimension. There are different methods to improve the crude truncations in the FRG literature. However, their applications for TGFTs remains difficult due to the non-locality of the interactions over the group manifold on which the fields are defined. A step to go out of the truncation method was done recently in \cite{Lahoche:2018oeo}-\cite{Lahoche:2018vun} with the \textit{effective vertex expansion} (EVE) method. Basically, the strategy is to close the infinite tower of equations coming from the exact flow equation, instead of crudely truncate them. To say more, the strategy is to complete the structure equation \eqref{structure} with a structure equation for $\Gamma^{(6)}$, expressing it in terms of the marginal coupling $\lambda$ and the effective propagator $G_s$ only. In this way, the flow equations around marginal couplings are completely closed. Note that this approach cross the first hypothesis motivating the truncation: We expect that so far from the deep UV, only the marginal interactions survive, and drag the complete RG flow. Moreover, any fixed point of the autonomous set of resulting equations are automatically fixed points for any higher effective melonic vertices building from effective quartic interactions. Finally, a strong improvement of this method with respect to the truncation method, already pointed out in \cite{Lahoche:2018oeo}-\cite{Lahoche:2018vun} is that it allows to keep the complete momenta dependence of the effective vertex. This dependence generate a new term on the right hand side of the equation for $\dot{Z}$, moving the critical line from its truncation's position. \\

\noindent
Let us consider the flow equation for $\dot{\Gamma}^{(2)}$, obtained from \eqref{Wetterich} deriving with respect to $M$ and $\bar{M}$:
\begin{equation}
\dot{\Gamma}^{(2)}(\vec{p}\,)=-\sum_{\vec{q}} \Gamma_{\vec{p},\vec{p},\vec{q},\vec{q}}^{(4)}\,G_s^2(\vec{q}\,) \dot{r}_s(\vec{q}\,)\,,\label{gamma2}
\end{equation}
where we discard all the odd contributions, vanishing in the symmetric phase, and took into account the corollary \ref{cor1}. Deriving on both sides with respect to $p_1^{2\alpha}$, and setting $\vec{p}=\vec{0}$, we get, in accordance with definition \ref{def1}:
\begin{equation}
\dot{Z}=-\sum_{\vec{q}} \Gamma_{\vec{0},\vec{0},\vec{q},\vec{q}}^{(4)\,\prime}\,G_s^2(\vec{q}\,)  \dot{r}_s(\vec{q}\,)-\Gamma_{\vec{0},\vec{0},\vec{q},\vec{q}}^{(4)}\,G_s^2(\vec{q}\,) \dot{r}_s(\vec{q}\,)\,,
\end{equation}
where the "prime" designates the partial derivative with respect to $p_1^{2\alpha}$. In the deep UV ($k\gg1$) the argument used in the $T^4$-truncation to discard non-melonic contributions holds, and we keep only the melonic diagrams as well. Moreover, to capture the momentum dependence of the effective melonic vertex $\Gamma_{\text{melo}}^{(4)}$ and compute the derivative $\Gamma_{\text{melo}\,,\vec{0},\vec{0},\vec{q},\vec{q}}^{(4)\,\prime}$, the knowledge of $\pi_{pp}$ is required. It can be deduced from the same strategy as for the derivation of the structure equation \eqref{structure}, up to the replacement :
\begin{equation}
\mathcal{A}_s \to \mathcal{A}_s(p):=\sum_{\vec{p}\in(\mathbb{Z}^D)^d}\,G^2_s(\vec{p}\,)\delta_{p_1p}\,,
\end{equation}
from which we get:
\begin{equation}
\pi_{pp}=\frac{2\lambda_r}{1+2\lambda_r\bar{\mathcal{A}}_s(p)}\,,\quad \bar{\mathcal{A}}_s(p):={\mathcal{A}}_s(p)-{\mathcal{A}}_{-\infty}(0)\,.
\end{equation}
The derivative with respect to $p_1^{2\alpha}$ may be easily performed, and from the renormalization condition \eqref{rencond}, we obtain: 
\begin{equation}
\pi_{00}^\prime=-4\lambda^2(s)\,\mathcal{A}_s^{\prime}\,,
\end{equation}
and the leading order flow equation for $\dot{Z}$ becomes:
\begin{equation}
\dot{Z}=4\lambda^2 \mathcal{A}_s^{\prime}(0) \,I_2(0)-2\lambda I_2^\prime(0)\,.
\end{equation}
As announced, a new term appears with respect to the truncated version \eqref{syst1}, which contains a dependence on $\eta$ and then move the critical line. The flow equation for mass may be obtained from \eqref{gamma2} setting $\vec{p}=\vec{0}$ on both sides. Finally, the flow equation for the marginal coupling $\lambda$ may be obtained from the equation \eqref{Wetterich} deriving it twice with respect to each means fields $M$ and $\bar{M}$. As explained before, it involves $\Gamma^{(6)}_{\text{melo}}$ at leading order, and to close the hierarchy, we use  the marginal coupling as a driving parameter, and express it in terms of $\Gamma_{\text{melo}}^{(4)}$ and $\Gamma^{(2)}_{\text{melo}}$ only. One again, from proposition \ref{propmelons}, $\Gamma^{(6)}_{\text{melo}}$ have to be split into $d$ monocolored components $\Gamma^{(6)\,,i}_{\text{melo}}$:
\begin{equation}
\Gamma^{(6)}_{\text{melo}}=\sum_{i=1}^d \Gamma^{(6)\,,i}_{\text{melo}}\,. 
\end{equation}
The structure equation for $ \Gamma^{(6)\,,i}_{\text{melo}}$ may be deduced following the same strategy as for  $\Gamma^{(4)\,,i}_{\text{melo}}$, from proposition \eqref{propmelons}. Starting from a vacuum diagram, a leading order $4$-point graph may be obtained opening successively two internal tadpole edges, both on the boundary of a common internal face. This internal face corresponds, for the resulting $4$-point diagram to the two external faces of the same colors running through the interior of the diagram. In the same way, a leading order $6$-point graph may be obtained cutting another tadpole edge on this resulting graph, once again on the boundary of one of these two external faces.
The reason this works is that,  in this may, the number of discarded internal faces is optimal, as well as the power counting. From this construction, it is not hard to see that the zero-momenta $ \Gamma^{(6)\,,i}_{\text{melo}}$ vertex function must have the following structure (see \cite{Lahoche:2018oeo}-\cite{Lahoche:2018vun} for more details):
\begin{equation}
 \Gamma^{(6)\,,i}_{\text{melo}}=(3!)^2\,\left(\vcenter{\hbox{\includegraphics[scale=0.5]{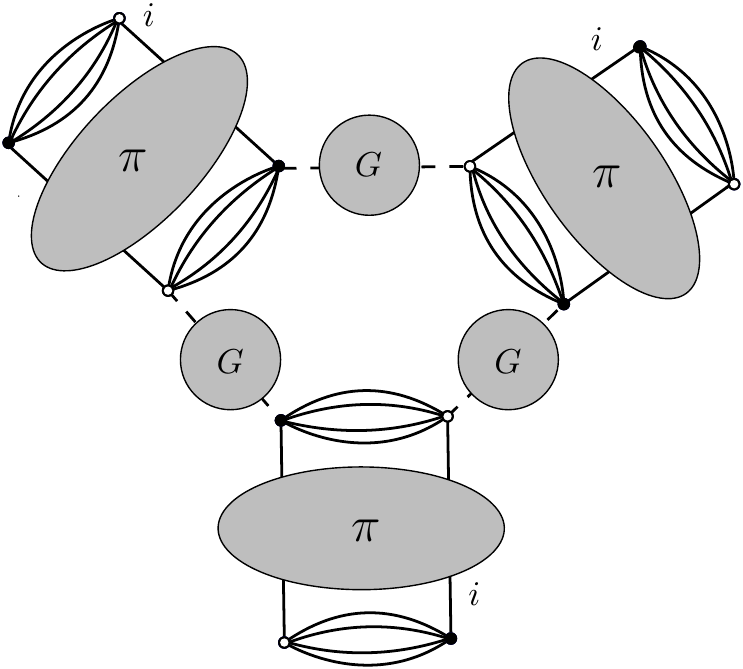} }}\right)\,,
\end{equation}
the combinatorial factor $(3!)^2$ coming from permutation of external edges. Translating the diagram into equation, and taking into account symmetry factors, we get:
\begin{equation}
 \Gamma^{(6)\,,i}_{\text{melo}} = 48Z^3(s)\bar{\lambda}^3(s)e^{-2\alpha s} \bar{\mathcal{A}}_{2s}\,,
\end{equation}
with:
\begin{equation}
\bar{\mathcal{A}}_{2s}:= Z^{-3}e^{2\alpha s} \sum_{\vec{p}\in(\mathbb{Z}^D)^{(d-1)}} G_s^3(\vec{p}\,)\,.
\end{equation} 
Note that this structure equation may be deduced directly from Ward identities, as pointed-out in \cite{Lahoche:2018vun} and \cite{Samary:2014tja}. The equation closing the hierarchy is then compatible with the constraint coming from unitary invariance. The flow equations involve now some new contributions depending on two sums, $\bar{\mathcal{A}}_{2s}$ and $\bar{\mathcal{A}}_s^{\prime}$, defined without regulation function $\dot{r}_s$. However, they are both power-counting convergent in the UV, and the renormalizability theorem ensures their finitness for all orders in the perturbation theory. For this reason, they becomes independent from the initial conditions at scale $\Lambda$ for $\Lambda\to\infty$; and as pointed out in \cite{Lahoche:2018vun}, it is suitable to use of the approximation \eqref{gamma2approx}. Explicitly, following the same computational strategy as for the calculation of $J_n(R)$ in section \ref{sec2}, we get, using the Litim's regulator:
\begin{equation}
\bar{\mathcal{A}}_{2s}=\frac{1}{2}\frac{\iota(d,D)}{1+\bar{m}^{2\alpha}}\left[\frac{1}{(1+\bar{m}^{2\alpha})^2}+\left(1+\frac{1}{1+\bar{m}^{2\alpha}}\right)\right]\,,
\end{equation}
and
\begin{equation}
\bar{\mathcal{A}}_s^{\prime}=\frac{1}{2}\iota(d,D)\frac{1}{1+\bar{m}^{2\alpha}}\left(1+\frac{1}{1+\bar{m}^{2\alpha}}\right)\,.
\end{equation}
The complete flow equation for zero-momenta $4$-point coupling write explicitly as:
\bea\label{flowfour}
\dot{\Gamma}^{(4)}=-\sum_{\vec p}\dot r_s(\vec p\,) G^2_s(\vec p\,)\Big[\Gamma^{(6)}_{\vec p,\vec 0,\vec 0,\vec p,\vec 0,\vec 0}
\quad-2\sum_{\vec p\,'}\Gamma^{(4)}_{\vec p,\vec 0,\vec p\,',\vec 0}G_s(\vec p\,')\Gamma^{(4)}_{\vec p\,',\vec 0,\vec p,\vec 0}+2 G_s(\vec p\,)[\Gamma^{(4)}_{\vec p,\vec 0,\vec p,\vec 0}]^2\Big].\cr
\eea
Keeping only the melonic contributions, we get finally the following autonomous system replacing the truncated flow equations \eqref{syst2} in the Litim's regulation:
\begin{align}
\left\{
    \begin{array}{ll}
       \beta_m&=-(2\alpha+\eta)\bar{m}^{2\alpha}-2\alpha d\bar{\lambda}\,\frac{\iota(d,D)}{(1+\bar{m}^{2\alpha})^2}\,\left(1+\frac{\eta}{6\alpha}\right)\,, \\
       \beta_{\lambda}&=-2\eta \bar{\lambda}+4\alpha\bar{\lambda}^2 \,\frac{\iota(d,D)}{(1+\bar{m}^{2\alpha})^3}\,\left(1+\frac{\eta}{6\alpha}\right)\Big[1-\iota(d,D)\bar{\lambda}\left(\frac{1}{(1+\bar{m}^{2\alpha})^2}+\left(1+\frac{1}{1+\bar{m}^{2\alpha}}\right)\right)\Big]\,. \label{syst3}
    \end{array}
\right.
\end{align}
where the anomalous dimension is then given by:

\begin{equation}
\eta=4\bar{\lambda}\iota(d,D)\frac{(1+\bar{m}^{2\alpha})^2-\frac{1}{2}\bar{\lambda}\iota(d,D)(2+\bar{m}^{2\alpha})}{(1+\bar{m}^{2\alpha})^2\Omega(\bar{\lambda},\bar{m}^{2\alpha})+\frac{(2+\bar{m}^{2\alpha})}{3}\bar{\lambda}^2[\iota(d,D)]^2}\,.
\end{equation}
The new anomalous dimension has two properties which distinguish him from its truncation version. First of all, as announced, the singularity line $\Omega=0$ moves toward the $\bar{\lambda}$ axis, extending the symmetric phase domain.
In fact, the improvement is \textit{maximal}, the critical line being deported under the singularity line $\bar{m}^{2\alpha}=-1$.  In standard interpretations \cite{Lahoche:2018oeo}, the presence of the region $II$ is generally assumed to come from a bad expansion of the effective average action around vanishing means field, becoming a spurious vacuum in this region.

\begin{figure}[H]\begin{center}
\includegraphics[scale=0.3]{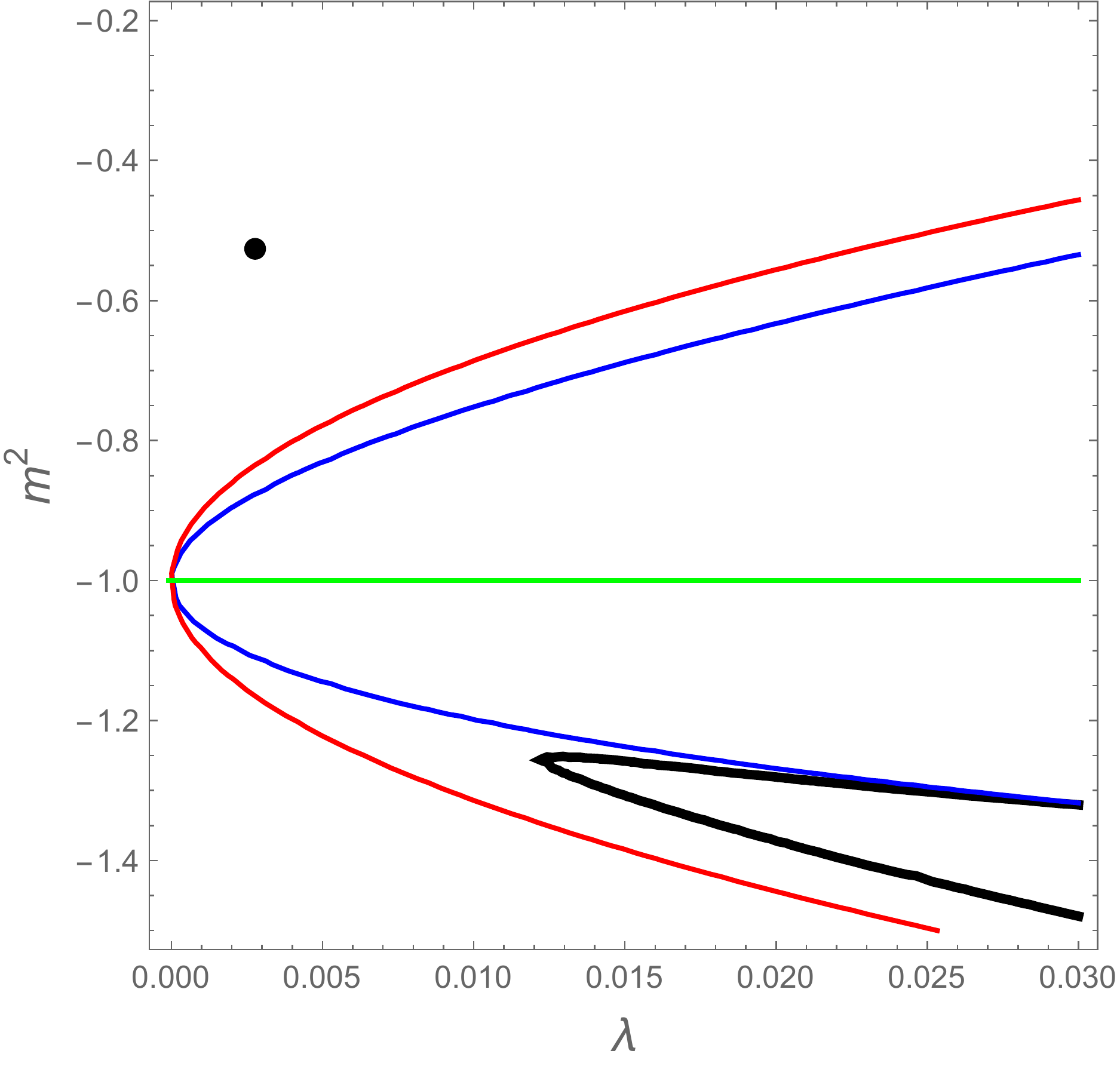} 
\caption{The relevant lines over the maximally extended region $I^\prime$, bounded at the bottom with the singularity line $m^{2}=-1$ (in green). The blue and red curves correspond respectively to the equations $L=0$ and $\Omega=0$. Moreover, the black point correspond to the numerical non-Gaussian fixed point, so far from the two previous physical curves. }\label{fig3}\end{center}
\end{figure}

However, the EVE method show that this singularity line is completely discarded taking into account the momentum dependence of the effective vertex. The second improvement come from the fact that the anomalous dimension may be negative, and vanish on the line of equation $L(\bar{\lambda},\bar{m}^{2\alpha})=0$, with:
\beq
L(\bar{\lambda},\bar{m}^{2\alpha}):=(1+\bar{m}^{2\alpha})^2-\frac{1}{2}\bar{\lambda}\iota(d,D)(2+\bar{m}^{2\alpha})\,.
\eeq
Figure \ref{fig3} summarize the analysis for $D=1$ and $d=5$. Interestingly, there are now two lines in the maximally extended region $I^\prime$ where physical fixed points are expected. However, numerical integrations for various kind of values $(D,d)$ show that the improved flow equations admit the non-Gaussian fixed points, from which one of them is numerically very close from the fixed point $p_+$ obtained in the truncation method, and then unphysical as well.  The other solutions are:
\bea
p_{0}=(\bar{m}^2=-1.28,\bar{\lambda}=0.025),\quad p_{1}=(\bar{m}^2=1.96,\bar{\lambda}=1.10),\quad 
\eea
For $p_0$ we have $\bar{m}^2<-1$. This fixed point cannot be taking into account by considering all the explanation given in the section \eqref{sec2}. $p_1$ have the following critical exponent $\theta_1=-2.8-4.2i$, $\theta_2=-2.8+4.2i$. This fixed point is IR attractive and lives in the same region like $p_+$. Finally all the fixed point discovered from EVE method violate the Ward identities.


\section{  Conclusion}\label{sec6}

In this paper we show that the IR fixed point obtained in the FRG applications for TGFT lack an important constraint coming from Ward identities. This constraint reduces the physical region of the phase space to a one-dimensional subspace without fixed point, suggesting that the phase transition scenario abundantly cited in the TGFT literature may be an artifact of an incomplete method. This suggestion is improved with a more sophisticated method, taking into account the momentum dependence of the effective vertex, and providing a maximal extension of the symmetric region. Despite with this improvement, the resulting numerical fixed point does not cross any of the physical lines provided from the Ward constraint. In the literature, the quartic truncation has been largely investigated, for various group manifold and dimensions. We expect from our analysis that none of these models modify our conclusions, except possibly for TGFT including \textit{closure constraint} as a Gauge symmetry. The definitive conclusion for this case remains a work in progress.



\end{document}